# Comparison of alpha- and beta- tin for lithium, sodium, and magnesium storage: an ab initio study including phonon contributions

F. Legrain,[1] O.I. Malyi,[2] C. Persson,[2,3] and S. Manzhos[1,a]

[1]*Department of Mechanical Engineering, National University of Singapore, 117576, Singapore*

[2]*Centre for Materials Science and Nanotechnology, University of Oslo, NO-0316 Oslo, Norway*

[3]*Department of Physics, University of Oslo, NO-0316 Oslo, Norway*

We present a comparative ab initio study of Li, Na, and Mg storage in tin, including phononic effects and phase competition between α and β Sn. Mg doping at low concentration is found to stabilize the β phase. On the contrary, Li and Na doping is shown to reverse the stability of the phases at room temperature: Li/Na-doped α-Sn is more stable than Li/Na-doped β-Sn up to a temperature of around 380/400 K. This may rationalize the formation of α-Sn upon lithiation and delithiation of β-Sn anodes reported in experimental studies. The changes in phase stability with Li/Na/Mg doping are directly related to the intercalation energies of Li/Na/Mg in one phase versus the other: at 300 K, Li/Na is easier intercalated in α-Sn (-0.37/-0.08 eV) than in β-Sn (0.06/0.49 eV), while Mg intercalation energy is, although positive (i.e. unfavored intercalation), lower in β-Sn (0.53 eV) than in α-Sn (0.66 eV). The temperature effect is found to affect significantly the intercalation energy, by up to 0.13 eV at 300 K. Analysis of diffusion barriers shows that Li, Na, and Mg diffusion in β-Sn is anisotropic with migration barriers along the (001) direction (respectively 0.01, 0.22, and 0.07 eV) significantly lower than those in α-Sn (respectively 0.20, 0.52, and 0.40 eV).

[a] Author to whom correspondence should be addressed. Electronic mail: mpemanzh@nus.edu.sg.

## I. INTRODUCTION

The development of more efficient electrochemical batteries is highly desired, driven in particular by the demand for ever smaller and lighter electronic devices, but also due to a widespread use of hybrid electrical vehicles and of renewable electricity - coming from clean but intermittent solar and wind energies.[1,2] Li-ion batteries are, as of today, the electrochemical battery technology providing the highest energy density and the only one widely commercialized.[3] However, Li resources are limited[4,5] and geographically localized.[6,7] There is a need to develop better Li-ion but also post-Li-ion batteries, such as Na- and Mg-ion batteries.[8-10] Among the main technological challenges remains the development of efficient host materials to (de)intercalate the metal atoms at the negative electrode.[8-10] We focus here on Sn, which has been substantially investigated experimentally as a promising anode material for all Li-, Na-, and Mg-ion batteries.[11-14] Sn provides good theoretical capacities: 994 mAh/g for Li, 847 mAh/g for Na, and 903 mAh/g for Mg, which have been approached experimentally.[11-14] Although Sn is a promising material, some of the very basic properties of Li, Na, and Mg storage in Sn remain unexplored and some of the phenomena observed during Li/Na/Mg (de)intercalation are still unexplained. In particular, while Sn is most stable at room temperature in the metallic β-Sn phase, the covalent α-Sn phase has been reported to form upon lithiation and delithiation of β-Sn based anodes, and eventually to become dominant after a number of cycles.[15-18] The mechanism of formation of the alpha phase during Li intercalation and de-intercalation in β-Sn based anodes is still not fully understood. The transition temperature between β-Sn and α-Sn is 286 K (13°C),[19] α (β)-Sn being more stable below (above) 286 K. Because the transition temperature is very close to the room temperature, the energy difference between the two phases at room temperature is tiny (about 0.001 eV per atom as computed here), and any external perturbation - such as Li/Na/Mg doping - can possibly lead to a phase competition between alpha and beta phases. Density functional theory (DFT) calculations investigating the effect of Li doping on the alpha-beta phase stability have been performed in previous studies.[18,20] However, sometimes opposite results were reported for DFT calculations. Im et al. show that Li doping stabilizes the beta phase,[18] whereas the results of Kaghazchi suggest that Li doping stabilizes the alpha phase.[20] Therefore, it remains unclear whether Li doping may be the cause of the formation of the α-Sn phase during lithiation and delithiation, and which phase of Sn is thermodynamically stable with Li doping. Moreover, there is to the best of our knowledge no single study on the effect of Na doping on the relative alpha-beta phase stability of Sn. However, the effect of Li, Na, and Mg doping on the relative stability of the two phases is critical because the phases present will determine the storage energetics (i.e. voltages) as well as the Li, Na, and Mg diffusion rate (i.e. charge/discharge rate) in battery practical electrode materials.

Previous calculations of Li (and Mg) insertion in Sn[18,20-23] ignored the vibrational contributions which are critical for phase stability and can also have a significant impact on storage properties. We also note that while Mg insertion in Sn has been studied by DFT,[21] the diffusion of Mg in β-Sn has not been considered along the (001) direction, while it is the one providing the lowest migration barrier for Li.[22] In addition, an ab initio study of Na insertion in β-Sn remains, to the best of our knowledge, absent from the literature: the insertion energetics as well as the migration barriers remain unknown.

Therefore, even the very basic and upstream questions remain not properly answered concerning Sn anodes: which of phases, α or β, is more stable with Li and Na doping, what is the insertion energetics of Na in β-Sn, what are the diffusion barriers of Na and Mg in β-Sn, and what is the effect of temperature on the alpha-beta phase stability with Li/Na/Mg doping? Given the typical DFT errors, to compare the effects due to the intercalation of different ions on both phases, it is important to perform calculations at the same level of theory and with the same computational setup. For example, for the migration barrier of Li in α-Sn, a value of 0.15 eV was reported in Ref. 20, a value of 0.24 in Ref. 24, and a value of 0.39 in Ref. 25, i.e. differences which are multiples of $k_BT$ (at room temperature) and would change the diffusion rate by orders of magnitude. Therefore, a truly comparative study is required.

Here, we endeavour to answer these questions in a systematic and first truly comparative ab initio study of all Li, Na, and Mg in α- and β-Sn. We study the relative stability of the Li/Na/Mg-doped α- and β-Sn as well as the intercalation energies of Li/Na/Mg in the two phases of Sn, including phononic contributions. We also investigate Li, Na, and Mg diffusion in α- and β-Sn, which is critical for understanding the charge/discharge rate of tin-based battery electrodes.

## II. METHODS

All calculations were performed using DFT with a plane basis set as implemented in the Vienna ab initio package (VASP)[26]. To describe electron-ion interactions, we applied the projector augmented wave (PAW) method.[27] The valence electrons considered were $2s^1$ for Li, $3s^1$ for Na, $3s^2$ for Mg, and $5s^2 5p^2$ for Sn. The effect of the addition of the 4d Sn electrons to the valence shell was found to be insignificant. To describe the electron exchange correlation, we applied the generalized gradient approximation using the Perdew-Burke-Ernzerhof (PBE) functional.[28] The plane wave basis was used with an energy cutoff of 300 eV. We have confirmed that the increase of the energy cut-off does not affect the results. We used 64-atom cells to model the α- and β-Sn phases, which correspond to 2×2×2 and 2×2×4 supercells, respectively. 4×4×4 and 8×8×8 Γ-centered Monkhorst-Pack meshes[29] were used for the *k*-point sampling of the Brillouin zone integration for α- and β-Sn, respectively. The different *k*-point sampling is used due to the metallic nature of β-Sn requiring a much denser *k*-point grid for energy convergence. Atomic coordinates and cell vectors were optimized until the forces on atoms were below 0.01 eV/Å. The lattice constants of α- and β-Sn were 6.65 and 5.93 Å, respectively, and the (electronic) energy difference between the phases 0.04 eV per atom, in good agreement with available data.[30] Diffusion barriers were computed with the nudged elastic band method.[31] To compute the migration energies in β-Sn, we first identified the transition state with constrained optimization by using 8 images, a 250 eV plane wave cutoff, and a 4×4×4 *k*-mesh for the Brillouin zone sampling. In a second step, we further recomputed the configuration found at the extremum of the energy curve (by interpolation) by using the very accurate setup described above (i.e. a 300 eV cutoff and a 8×8×8 *k*-mesh for the Brillouin zone sampling). Yet, for Mg diffusion along the (001) direction in β-Sn, a lowering in energy (of around 0.03 eV) was found at the transition state due to reorganization of some Sn atoms because of inter-cell dopant-dopant interaction along the (100) direction. The transition site was thus re-optimized by fixing along the (001) direction two Sn atoms located at 4.9 Å from the Mg transition site. In α-Sn, we directly optimized the transition state with the 300 eV cutoff and a 4×4×4 *k*-mesh, the transition site being known in the diamond structure.[20,21,32] To investigate the temperature effect, we used density functional perturbation theory (DFPT)[33]. Because of computational cost of DFPT calculations, we reduced the *k*-mesh sampling for β-Sn from 8×8×8 to 4×4×4. The phonon frequencies were obtained by using the PHONOPY code[34] together with a 40×40×40 *k*-mesh sampling. We applied the harmonic approximation, but not without first confirming that very similar results are obtained with the quasi-harmonic approximation.

The effect due to the zero-point energy (ZPE) and to finite temperature was considered by computing the contributions arising from the vibrational energy and entropy[30]:

$$E_{vib} - TS_{vib} = \int d\mathbf{k} \sum_{i=1}^{N} \left\{ \frac{1}{2} v_i + k_B T ln(1 - e^{-v_i/k_B T}) \right\}, \quad (1)$$

where $v_i$ is the energy of one quantum in the $i^{th}$ normal mode, $k_B$ is be Boltzmann constant, $\mathbf{k}$ is the wave vector, T is the temperature, and $S_{vib}$ is the vibrational entropy. Adding these vibrational contributions to the DFT ground state energy $E_{DFT}$ gives the Helmholtz free energy of the system *F*:

$$F = E_{DFT} + E_{vib} - TS_{vib}. \quad (2)$$

To reproduce the transition temperature between α- and β-Sn (~290 K), which is highly dependent on the difference in DFT ground state energies between the two phases ($E_{DFT}^{beta}$-$E_{DFT}^{alpha}$), we applied a correction of 0.013 eV per atom to the β-Sn's $E_{DFT}$.

The energetics of Li, Na, and Mg insertion in alpha- and beta-Sn were analyzed based on the defect formation energies (intercalation energies) $E_f$:

$$E_f = \left( E(Sn-M) - E(Sn) - n \times E(M) \right)/n, \quad (3)$$

where M stands for the inserted metal (M=Li/Na/Mg), *n* is the number of metal atoms inserted, *E(Sn-M)* designates the energy of the Li/Na/Mg-inserted Sn structures, *E(Sn)* represents the energy of the pure Sn phases (α- and β-Sn), and *E(M)* represents the energy of one atom of Li/Na/Mg in bulk (i.e. *bcc* Li/Na and *hcp* Mg). The metal reference states of Li, Na, and Mg were computed using unit cells together with a 500 eV plane wave cutoff and a 20×20×20 *k*-mesh sampling for the Brillouin zone. The phonon calculations were carried out on supercells of size of about 10×10×10 Å with a 300 eV plane wave cutoff and 6×6×6 (8×8×8) *k*-mesh for Li/Na (Mg).

## III. RESULTS AND DISCUSSION

### A. Insertion sites

In the diamond structure of α-Sn, Li, Na, and Mg insert at the tetrahedral site (see Fig. 1(a)), similarly to Li insertion in Si.[32] The Li/Mg/Na-Sn bonds in α-Sn are found to be 2.96/3.00/3.03 Å (3.34/3.35/3.36 Å) for the first (second) coordination sphere, in agreement with previous studies.[20,21] In β-Sn, the unit cell was screened to identify the insertion sites of Li, Na, and Mg. The screening was done using a grid with a resolution of ~0.75 Å along the (100) and (010) directions and of ~0.8 Å along the (001) direction. Points within a 1.8 Å from a Sn atom were removed, and all other points were tested as possible insertion sites. For Li and Mg, two very close kinds of sites (separated by 0.15 and 0.09 Å, respectively) with similar energetics (within 0.0035 eV for Li and 0.005 eV for Mg) are found. The most symmetric site (shown in Fig. 1) is sevenfold-coordinated, and is located between two equivalent sites of the other kind, which are displaced from the first site along the $c$ axis by +/-0.012$c$ for Li and by ±0.006$c$ for Mg (c being the length of the lattice vector of the 64 atom cell). As explained below, the transition sites for Li/Mg diffusion along the c direction give very similar energetics to these first two sites too (within 0.01 eV for Li, and of 0.07 eV for Mg). Rather than being located at a single well-defined site, Li (and to a certain extent Mg) are expected to be distributed along the $c$ channel (displayed in Fig. 2). This is in contrast to α-Sn, where insertion sites are well defined, i.e. they are deep local minima. The Li/Mg-Sn closest bond is 2.67-2.68/2.84-2.85 Å for the sites identified along the $c$ channel for Li/Mg. The lowest insertion site for Na is at a similar position to the so-called most symmetric sites of Li and Mg, but Na displaces one Sn atom by 1.8 Å, resulting in a threefold-coordinated insertion site and Na-Sn shortest bonds of 3.09 Å.

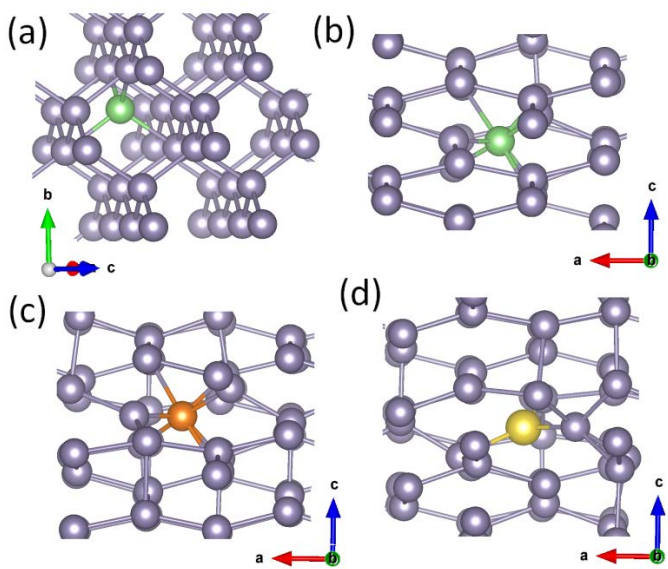

**Figure 1**. Insertion sites of (a) Li in α−Sn (the same sites hold for Na and Mg); (b) Li in β-Sn; (c) Mg in β-Sn; (d) Na in β-Sn. Colors here and elsewhere: green – Li, orange – Mg, yellow – Na, grey – Sn.

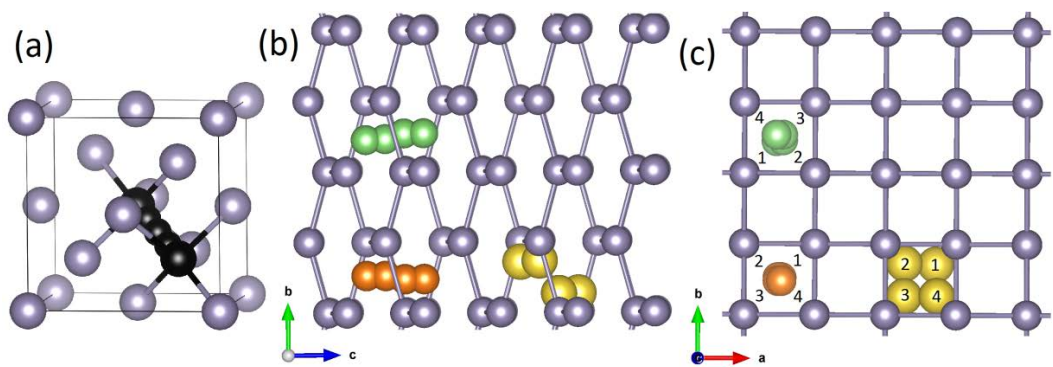

**Figure 2**. (a) Migration pathways for Li/Na/Mg (in black) in α-Sn. The larger black atoms show the equilibrium sites for Li/Na/Mg in α-Sn while the smaller black atoms depict the migration pathways between two equilibrium sites; (b) and (c) Migration pathways for Li (green), Na (yellow), and Mg (orange) along the (001) direction in β-Sn (grey). (b) and (c) show the (100) and (001) planes, respectively.

**B. Insertion energetics**

The defect formation energies of Li, Na, and Mg in the two phases of Sn are given in Table I. As the defect formation energies are computed versus the metal reference states of Li/Na/Mg, positive (negative) values indicate that the insertion of a single Li/Na/Mg dopant in Sn is unfavored (favored) versus the formation of Li/Na/Mg bulk metal. For β-Sn, which is the most stable phase of Sn at room temperature, all defect formation energies are found to be positive. Hence, the insertion of all single Li, Na, and Mg dopants is unfavored in β-Sn. The positive values for (i.e. unfavored) single dopant insertion computed here may explain the front between a doped and an undoped phase which was observed upon Na insertion in β-Sn.[35] Even though all defect formation energies are positive, the value is significantly lower for Li than for Na/Mg, suggesting that Li insertion in β-Sn is easier than that of Na/Mg. In α-Sn however, the defect formation energies for Li and Na insertion are lower than in β-Sn and negative. This indicates that single Li/Na dopant insertion in α-Sn is both easier than in β-Sn and favored versus Li/Na metal formation. For Mg, however, the defect formation energy is even higher in α than in β-Sn, suggesting a thermodynamically unfavored single Mg atom insertion in both phases. The highly positive $E_f$ for Mg in both phases of Sn suggests that the Sn-based anodes reported for Mg-ion batteries[14] do not operate by single atom Mg insertion and diffusion. Therefore, interfacial reactions and second phase formation might be necessary for Mg storage in Sn, possibly imposing the low charge/discharge rate reported.[14] The $E_f$ of Na insertion in β-Sn is also highly positive, suggesting that during the first cycles (and as long as there is no α-phase formed), the same issues as described above for Mg may be expected.

Insertion of multiple Li/Na/Mg atoms (i.e. 2, 4, and 8 dopants) was also performed in the 64-atom cells of α and β-Sn. Among the many possible configurations, we considered the configurations maximizing inter-distance dopants. We confirmed that well-dispersed configurations are indeed the lowest energy configurations in α-Sn. They are however not the lowest energy configurations in β-Sn. While computing all configurations for all concentrations is impractical, we considered a dozen configurations for the insertion of 2 Li/Na/Mg atoms, to evaluate the difference in energy that can exist between the well-dispersed configurations and the lowest energy configurations. The defect formation energies computed are given in Table II. At the concentration corresponding to the insertion of 2 dopants, the differences in $E_f$ between the well-dispersed configuration and the lowest energy configuration (among those computed) are inferior to 0.15 eV, indicating that the energies computed, although not necessarily the lowest, are reasonable. The results obtained for multiple atoms insertion show that for most systems, the defect formation energies computed for single dopants do not change significantly when the number of dopants per supercell is increased up to 8 (which corresponds to a capacity of around 30 (55) mAh/g for Li/Na (Mg)). Larger changes are observed for Mg insertion, but $E_f$ at all considered Mg concentrations remain significantly positive (unfavored insertion). The results of the present study are therefore relevant for a range of Li/Na/Mg concentrations at low state of charge.

To understand the difference in insertion energetics among different dopant types (Li/Na/Mg) and the host phases ($\alpha$-/$\beta$-Sn), we computed (i) the charge donated per dopant atom to the Sn framework, by using the Bader[36] method, and (ii) the so-called strain energies[37] computed as follows:

$$E_{strain} = E_{[64Sn+M,relaxed]-M} - E_{64Sn}, \qquad (4)$$

where $E_{[64Sn+M,relaxed]-M}$ is the energy of the Sn framework when accommodated to host a dopant atom, and $E_{64Sn}$ is the energy of the ideal Sn structure. The Bader charges and the strain energies are given in Table I. The Bader charges are generally used to assess the strength of the ionic interaction between the dopant atom and the host structure; the larger the donated charge (from the dopant atom to the host) the stronger the interaction. Our results show that, indeed, the order of the donated charges between the phases fits relatively well with that of the defect formation energies, even though the differences in Bader charges between $\alpha$-Sn and $\beta$-Sn remain small (even negligible for Mg). The strain energies can be used to assess the energy cost of dopant insertion because of the stress generated on the host structure to accommodate the dopant. Higher strain energies are computed for Na and Mg compared to Li, likely associated to the larger size of Na and bivalency of Mg. The results show that these higher strain energies for Na and Mg correspond to higher defect formation energies (i.e. weaker intercalation energies) versus that of Li.

**Table I**. Charge donation $q$ (in units of $|e|$) from Li/Na/Mg to the Sn framework, defect formation energies $E_f$ (in eV) and strain energies $E_{strain}$ (in eV) of Li, Na, and Mg insertion in α-Sn and β-Sn. For $E_f$, zero corresponds to the cohesive energy of Li, Na, and Mg, respectively.

| $E_f$ (eV) | Li | Na | Mg |
|---|---|---|---|
| α-Sn | -0.32 | -0.09 | 0.79 |
| β-Sn | 0.02 | 0.50 | 0.55 |
| $E_{strain}$ (eV) | Li | Na | Mg |
| α-Sn | 0.06 | 0.21 | 0.17 |
| β-Sn | 0.61 | 1.12 | 1.28 |
| $q$ (e) | Li | Na | Mg |
| α-Sn | 0.85 | 0.75 | 1.36 |
| β-Sn | 0.80 | 0.74 | 1.36 |

**Table II**. Defect formation energies (in eV) for the insertion of multiple dopants (1, 2, 4, 8) in the 64-atom cells of α and β-Sn. Well-dispersed configurations are used. For 2 dopants, a dozen of different configurations are considered and the lowest $E_f$ is also given, denoted by $2_{lowest}$.

| $E_f$ (eV) | Dopants | | Li | | Na | | Mg | |
|---|---|---|---|---|---|---|---|---|
| α-Sn | 1 | | -0.32 | | -0.09 | | 0.79 | |
|  | 2 | | -0.27 | | -0.06 | | 0.83 | |
|  | 4 | | -0.19 | | 0.05 | | 0.82 | |
|  | 8 | | -0.11 | | 0.07 | | 0.70 | |
| β-Sn | 1 | | 0.02 | | 0.50 | | 0.55 | |
|  | 2 | $2_{lowest}$ | 0.00 | -0.08 | 0.36 | 0.26 | 0.52 | 0.38 |
|  | 4 | | -0.07 | | 0.15 | | 0.33 | |
|  | 8 | | -0.017 | | 0.34 | | 0.96 | |

**C. Phononic effects**

The energy difference between α- and β-Sn at room temperature is computed to be as tiny as 0.001 eV per atom. The temperature dependence of the Helmholtz free energies for pure and Li/Na/Mg-doped Sn is plotted in Fig. 3. The Helmholtz free energies are offset by the value of (pure or doped) α-Sn at 0 K in order to highlight the effect due to finite temperature on the phase stability between α- and β-Sn. The comparison between the plots shows that Li and Na doping stabilizes the alpha phase at room temperature. The alpha phase is stable up to 380 K for Li and 400 K for Na (for x=1/64, x being the number of Li/Na dopant atoms per Sn atom). On the contrary, Mg doping stabilizes the β-Sn phase; it reduces the transition temperature by ~30 K (to ~260 K). This is a direct consequence of the lower (higher) $E_f$ for the insertion of Li/Na (Mg) in α-Sn versus β-Sn. The stabilization of α-Sn with Li doping could rationalize the formation of α-Sn upon lithiation reported in experimental studies.[15-17] For Mg-ion batteries, to the best of our knowledge, the stabilization of α-Sn during the cycling was not observed experimentally,[14] which is consistent with our results.

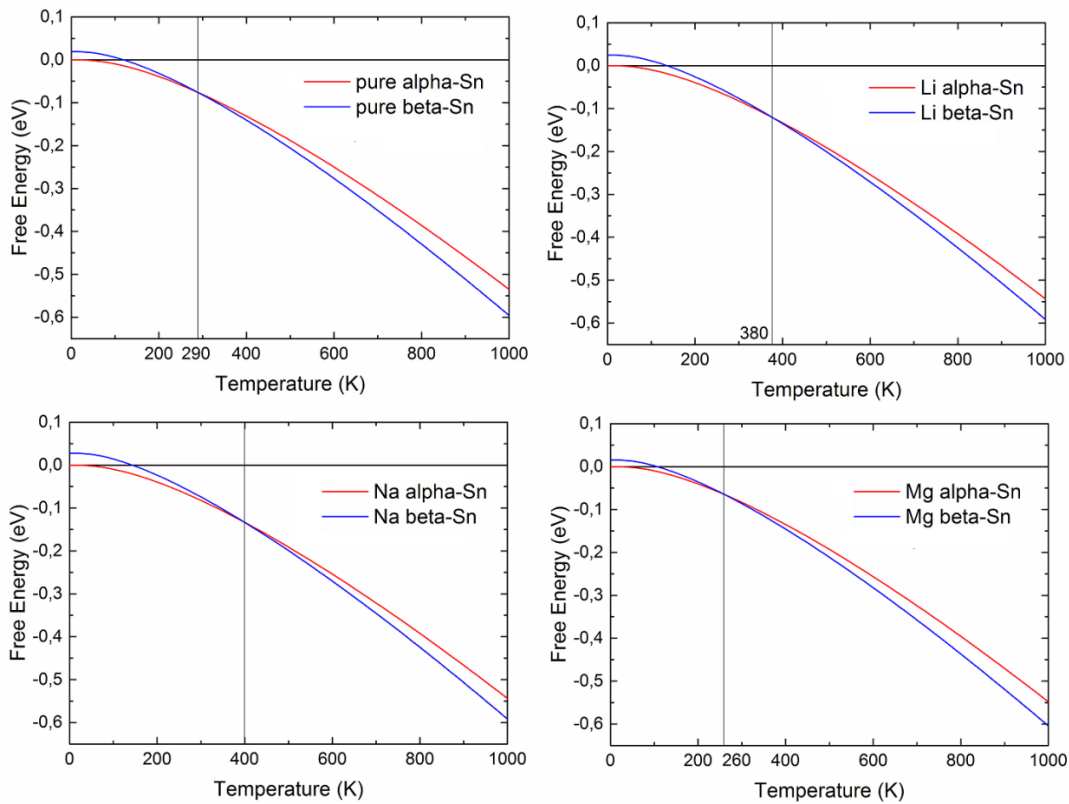

**Figure 3**. Helmholtz free energies - offset by the value of (pure or doped) α-Sn at 0 K - against temperature for pure and Li/Na/Mg-doped Sn (in eV). The vertical lines indicate the abscises of the intersection points.

In Table III, we summarize the effect of temperature (for $T = 300$ K) on the defect formation energies for the insertion of Li/Na/Mg in α- and β-Sn. The change in the energies due to $E_{vib}$-$TS_{vib}$ is found to be consistent with a previous study restricted to α-Sn[24] which used a very different computational setup (atom-centered bases and norm-conserving pseudopotentials[38]). The results show that this effect due to the zero point energy and to finite temperature is not always negligible (i.e. does not always cancel out in energy differences): it is as high as 0.13 eV for Mg in α-Sn (cf. to $k_BT = 0.026$ eV at 300 K).

**Table III**. Temperature effect at $T=300$ K on the defect formation energies Ef for Li/Na/Mg insertion in α- and β-Sn: the defect formation energies without considering the temperature effect ($E_{DFT}$ only), the temperature effect for $T=300$ K ($E_{vib}$-$TS_{vib}$ only), and the defect formation energies including the temperature effects ($F=E_{DFT}+E_{vib}$-$TS_{vib}$). Zero corresponds to the cohesive energy of Li, Na, and Mg, respectively. The values are in eV.

| host | $T=300$ K | Li | Na | Mg |
|---|---|---|---|---|
| α-Sn | $E_f$, w/o $T$ (i.e. $E_{DFT}$) | -0.32 | -0.09 | 0.79 |
| | $T$ effect (i.e. $E_{vib}$-$TS_{vib}$) | -0.05 | 0.01 | -0.13 |
| | $E_f$, w/ $T$ (i.e. $F$) | -0.37 | -0.08 | 0.66 |
| β-Sn | $E_f$, w/o $T$ (i.e. $E_{DFT}$) | 0.02 | 0.50 | 0.55 |
| | $T$ effect (i.e. $E_{vib}$-$TS_{vib}$) | 0.04 | -0.01 | -0.02 |
| | $E_f$, w/ $T$ (i.e. $F$) | 0.06 | 0.49 | 0.53 |

**D. Migration barriers**

In β-Sn, we investigated the (100) and (001) directions as potential migration pathways for Li, Na, and Mg (note that the (010) direction is equivalent to (100)). In the (100) direction, the migration barriers were found to be of 0.32 eV for Li, 0.35 eV for Mg, and 0.71 eV for Na. In the (001) direction, the migration barriers were found to be exceptionally small for Li: inferior to 0.01 eV, and relatively small for Mg (0.07 eV) and Na (0.22 eV). In the (001) direction, Li, Na, and Mg diffuses through a helix shape pathway, as presented in Fig. 2 and reported previously for Li. The points shown in Fig. 2 for the pathways in β-Sn are sites equivalent to those represented in Fig. 1 (given the symmetry of β-Sn). The Fig. 2 also shows that the helical shape of the pathway is more pronounced for Na than for Li and Mg.

The migration paths of Li, Na, and Mg in α-Sn are identical to the migration path of Li in Si[32]: diffusion happens between two tetrahedral sites via a hexagonal site. The barriers for Li, Na, and Mg in α- and β-Sn are given in Table IV. These results imply that Li, Na, Mg diffusion in beta Sn is anisotropic and with lowest migration barriers significantly lower than those in alpha Sn at the low metal concentration considered here.

**Table IV**. Migration barriers of Li, Na, and Mg in α-Sn and β-Sn (in eV).

| host | Li | Na | Mg |
|---|---|---|---|
| α-Sn | 0.20 | 0.52 | 0.40 |
| β-Sn (001) | <0.01 | 0.22 | 0.07 |
| β-Sn (100) | 0.32 | 0.71 | 0.35 |

IV. CONCLUSIONS

The effect of Li, Na, and Mg doping on the phase stability between the α- and β-Sn is investigated in this study. β-Sn is computed to be more stable than α-Sn at room temperature by only 0.001 eV per atom. It is shown that Li and Na doping (at x=1/64, x being the number of Li/Na dopants per Sn atom) reverses the phase stability between the alpha and beta phases at room temperature. This may explain the formation of α-Sn reported in experiments upon lithiation and delithiation. Mg doping, on the opposite, stabilizes β-Sn at room temperature. The stabilization of α/β-Sn with metal doping is directly related to the Li/Na/Mg intercalation energies, lower in beta (alpha) for Li/Na (Mg) than in α-Sn (β-Sn). This also indicates that Li and Na insertion are easier in α-Sn (- 0.37 and -0.08 eV) than in β-Sn (0.06 eV and 0.49 eV) while Mg insertion is easier in β-Sn (0.53 eV) than in α-Sn (0.66 eV).

The study also shows that the diffusion in β-Sn is anisotropic with the lowest migration barriers exceptionally small for Li ($<0.01$ eV) and relatively small for Mg (0.07 eV) and Na (0.22 eV), all significantly lower than those in α-Sn: 0.20 eV for Li, 0.40 eV for Mg, and 0.52 eV for Na.

The temperature effect on the insertion energetics was found to be as high as 0.13 eV for Mg in α-Sn, indicating that one should be cautious when neglecting the contributions arising from the vibrational energy and entropy.

## ACKNOWLEDGMENTS

This work was supported by the Ministry of Education of Singapore via an AcRF grant (R-265-000-494-112) and Research Council of Norway (contract No. 221469). This work was performed on the Abel Cluster, owned by the University of Oslo and the Norwegian Metacenter for Computational Science (NOTUR), and operated by the Department for Research Computing at USIT, the University of Oslo IT-department. F.L. is grateful to the University of Oslo for hospitality and financial support.

## REFERENCES

1  J. Rugolo and M. J. Aziz, Energy Environ. Sci. **5**, 7151 (2012).

2  V. Palomares, P. Serras, I. Villaluenga, K. B. Hueso, J. Carretero-Gonzalez and T. Rojo, Energy Environ. Sci. **5**, 5884 (2012).

3  Y. Tang, Y. Zhang, W. Li, B. Ma and X. Chen, Chem. Soc. Rev. **44**, 5926 (2015).

4  J. M. Tarascon, Nat. Chem. **2**, 510 (2010).

5  M. D. Slater, D. Kim, E. Lee and C. S. Johnson, Adv. Funct. Mater. **23**, 947 (2013).

6  A. K. Shukla and T. P. Kumar, J. Phys. Chem. Lett. **4**, 551 (2013).

7  C. Grosjean, P. H. Miranda, M. Perrin and P. Poggi, Renew. Sust. Energ. Rev. **16**, 1735 (2012).

8  M. N. Obrovac and V. L. Chevrier, Chem. Rev. **114**, 11444 (2014).

9  N. Yabuuchi, K. Kubota, M. Dahbi and S. Komaba, Chem. Rev. **114**, 11636 (2014).

10 J. Muldoon, C. B. Bucur and T. Gregory, Chem. Rev. **114**, 11683 (2014).

11 L. D. Ellis, P. P. Ferguson and M. N. Obrovac, J. Electrochem. Soc. **160**, A869 (2013).

12 L. D. Ellis, T. D. Hatchard and M. N. Obrovac, J. Electrochem. Soc. **159**, A1801 (2012).

13 Y. Wang, M. Wu, Z. Jiao and J. Y. Lee, Chem. Mater. **21**, 3210-3215 (2009).

14 N. Singh, T. S. Arthur, C. Ling, M. Matsui and F. Mizuno, Chem. Commun. **49**, 149 (2013).

15 L. Xu, C. Kim, A. K. Shukla, A. Dong, T. M. Mattox, D. J. Milliron and J. Cabana, Nano Lett. **13**, 1800 (2013).

16 H. Kim, Y. J. Kim, D. G. Kim, H. J. Sohn and T. Kang, Solid State Ionics **144**, 41 (2001).

17 K. Hirai, T. Ichitsubo, T. Uda, A. Miyazaki, S. Yagi and E. Matsubara, Acta Mater. **56**, 1539 (2008).

18 H. S. Im, Y. J. Cho, Y. R. Lim, C. S. Jung, D. M. Jang, J. Park, F. Shojaei and H. S. Kang, ACS Nano **7**, 11103 (2013).

19 M. J. Musgrave, Proc. Roy. Soc. (A) **272**, 503 (1963).

20 P. Kaghazchi, J. Phys.: Condens. Matter **25**, 382204 (2013).

21 Z. Wang, Q. Su, J. Shi, H. Deng, G. Q. Yin, J. Guan, M. P. Wu, Y. L. Zhou, H. L. Lou and Y. Q. Fu, ACS Appl. Mater. Inter. **6**, 6786 (2014).

22 Q. Li, P. Wang, Q. Feng, M. Mao, J. Liu, S. X. Mao and H. Wang, Chem. Mater. **26**, 4102 (2014).

23 J. Shi, W. Shi, W. Jin and G. Yin, Int. J. Electrochem. Sci. **10**, 4793 (2015).

24 F. Legrain, O. I. Malyi and S. Manzhos, in *Symposium N – Diamond Research Frontiers on Electrochemical Energy Storage Materials – Design, Synthesis, Characterization and Modeling*, San Francisco, USA, April 21-25 2014, edited by C. Wang (MRS Proceedings, 2014) **1678**, DOI: 10.1557/opl.2014.743.

25 C.-Y. Chou, M. Lee and G. S. Hwang, J. Phys. Chem. C **119**, 14843 (2015).

26 G. Kresse and J. Furthmüller, Comput. Mater. Sci. **6**, 15 (1996).

27 G. Kresse and D. Joubert, Phys. Rev. B **59**, 1758 (1999).

28 J. P. Perdew, K. Burke and M. Ernzerhof, Phys. Rev. Lett. **77**, 3865 (1996).

29 H. J. Monkhorst and J. D. Pack, Phys. Rev. B **13**, 5188 (1976).

30 S.-H. Na and C.-H. Park, J. Korean Phys. Soc. **56**, 494 (2010).

31 G. Henkelman, B. P. Uberuaga and H. Jónsson, J. Chem. Phys. **113**, 9901 (2000).

32 W. H. Wan, Q. F. Zhang, Y. Cui and E. G. Wang, J. Phys. Condens. Matter **22**, 415501 (2010).

33 S. Baroni, S. de Gironcoli, A. Dal Corso and P. Giannozzi, Rev. Mod. Phys. **73**, 515 (2001).

34 A. Togo, F. Oba and I. Tanaka, Phys. Rev. B **78**, 134106 (2008).

35 J. W. Wang, X. H. Liu, S. X. Mao and J. Y. Huang, Nano Lett. **12**, 5897 (2012).

36 W. Tang, E. Sanville and G. Henkelman, J. Phys. Condens. Matter **21**, 084204 (2009).

37 B. R. Long, M. K. Y. Chan, J. P. Greeley and A. A. Gewirth, J. Phys. Chem. C **115**, 18916 (2011).

38 M. S. José, A. Emilio, D. G. Julian, G. Alberto, J. Javier, O. Pablo and S.-P. Daniel, J. Phys. Condens. Matter **14**, 2745 (2002).